\documentstyle[11pt,aaspp4]{article}

\received{}
\accepted{}
\journalid{}{}
\articleid{}{}

\slugcomment{Accepted by {\it The Astrophysical Journal}}

\lefthead{Halpern \& Eracleous}
\righthead{No Binary Black Hole in OX 169}

\begin{document}

\rightskip 0pt \pretolerance=100

\title{The End of the Lines for OX 169: No Binary Broad-Line Region}

\author{J. P. Halpern\altaffilmark{1}}
\affil{Astronomy Department, Columbia University, 550 West 120th Street,
    New York, NY 10027}

\and

\author{M. Eracleous\altaffilmark{1}}
\affil{Department of Astronomy and Astrophysics, The Pennsylvania State
University, \\525 Davey Laboratory, University Park, PA 16802}

\altaffiltext{1}{Visiting Astronomer, Kitt Peak National Observatory,
National Optical Astronomy Observatories, which is operated by the
Association of Universities for Research in Astronomy, Inc. (AURA)
under cooperative agreement with the National Science Foundation.} 

\begin{abstract}
\rightskip 0pt \pretolerance=100 \noindent
We show that unusual Balmer emission line profiles of the quasar OX~169,
frequently described as either self-absorbed or double peaked,
are actually neither.  The effect is an illusion resulting
from two coincidences.  First, the forbidden lines
are quite strong and broad.  Consequently, the
[N~II]~$\lambda$6583 line and the associated narrow-line
component of H$\alpha$ present the appearance of twin 
H$\alpha$ peaks.  Second, the redshift of 0.2110 brings
H$\beta$ into coincidence with Na~I~D at zero redshift,
and ISM absorption in Na~I~D divides the
H$\beta$ emission line.  In spectra obtained over the past decade,
we see no substantial change in the character of the line profiles,
and no indication of intrinsic double-peaked structure. 
The H$\gamma$, Mg~II, and Ly$\alpha$ emission lines
are single peaked, and all of the emission-line
redshifts are consistent once they are correctly
attributed to their permitted and forbidden-line identifications.
A systematic shift of up to 700~km~s$^{-1}$ between broad
and narrow lines is seen, but such differences are common,
and could be due to gravitational and transverse redshift
in a low-inclination disk.
Stockton \& Farnham (1991) had called attention
to an apparent tidal tail in the host galaxy of OX~169,
and speculated that a recent
merger had supplied the nucleus with a coalescing pair of black holes
which was now revealing its existence in the form of two physically
distinct broad-line regions.  Although there is no longer 
any evidence for two broad emission-line regions
in OX 169, binary black holes should form frequently in galaxy mergers,
and it is still worthwhile to monitor the radial velocities of
emission lines which could supply
evidence of their existence in certain objects.
\end{abstract}

\section{Introduction}

The quasar OX~169 is a compact radio source (Gower \& Hutchings 1984)
as well as a source of curious optical emission lines, the
nature of which has been
the subject of interesting speculation for 20 years.  Smith
(1980) first noted the presence of apparent ``self-absorption'' in the
Balmer lines of OX~169 which, if correct as a physical description,
would be highly unusual for non-resonance lines.  Gaskell (1981)
preferred an alternative explanation based on displaced velocities
in which the broad and narrow-line regions differ
by 1200~km~s$^{-1}$.  Ten years later, Stockton and Farnham
(1991, hereafter SF)
interpreted variability of the H$\beta$ line profile as evidence for
two distinct {\it broad} peaks, thus assigning OX~169
to the family of double-peaked emitters (e.g., Chen \& Halpern 1989;
Eracleous \& Halpern 1994) whose origin remains a subject of intense study.
SF discussed both accretion-disk and binary broad-line region (BLR)
models, but settled on the binary explanation
as more consistent with the nature of the
variability (actually, only the difference between two spectra), and
the one in their estimation to be the most likely to account for the
range of double-peaked behavior seen up to that time
in active galaxies in general.
Of considerable interest was the connection made by SF between their
emission-line evidence for a binary black hole, and
an apparent tidal tail in the host galaxy which they showed had
the spectrum of starlight.   To the extent that black
holes are thought to be common in galactic nuclei, and in view of the
appearance of OX~169 as a recent merger, SF speculated that
a bound pair of black holes had formed, and was now
revealing its existence in the form of two distinct BLRs.
In this interpretation, it was {\it assumed} that a pair of supermassive black
holes can maintain physically distinct BLRs that appear to the observer
sufficiently separated in velocity.

In this paper, we re-evaluate these interesting suggestions about
OX~169 using an extensive set of optical spectra obtained over the past
decade, as well as archival ultraviolet spectra from the
{\it Hubble Space Telescope} ({\it HST\/}).  Our conclusion is that
there is little if any spectroscopic evidence for a binary BLR in OX~169.
In addition to revising the observational
description of the broad emission lines from double peaked to single
peaked,
we discuss how line-profile variability figured into previous
interpretations, how our understanding of line-profile variability
has developed over the past decade, and what lines of investigation
remain to be pursued in this subject.

\section{Observations}

We obtained many spectra of OX~169 over the past 10 years, with
resolution in the range 4--12 \AA.  Most of these covered the H$\beta$
and H$\alpha$ emission lines only, although H$\gamma$ and Mg~II were
each observed once.  A log of the spectroscopic observations is given
in Table~1.  Reductions were performed using standard techniques.  Of
particular relevance to this study is the accuracy of the wavelength
calibration, which is typically better than 1/20th of the resolution
as determined by the dispersion in the fits to the arc lines.
Wavelength calibrations were taken immediately after each object
exposure.  Slit widths were in the range $1.\!^{\prime\prime}7 -
2.\!^{\prime\prime}0$, and depending upon seeing and guiding,
placement of the object within the slit can be the dominant source of
systematic error in wavelength.  The dispersion among measurements of
the stronger lines in different spectra of OX 169 is approximately
1~\AA, which we interpret as a systematic uncertainty of $\approx
50$~km~s$^{-1}$ in velocity.  Our observational technique was not
designed to achieve photometric precision in flux.  Instead, we rely on
assumed constancy of the
[O~III]~$\lambda$5007 line wherever possible to standardize the flux.

We also make use of archival
({\it HST\/}) spectra of OX~169 obtained in 1992 with the
Faint Object Spectrograph (FOS) gratings G130H, G190H, and G270H. 
These were initially reported by Diplas et al. (1993).
Details of the FOS spectra are also given in Table~1.

\subsection{Narrow Emission Lines}

We begin by assessing the ``systemic'' redshift of OX~169 as defined
by its low-ionization forbidden lines and narrow components of its
Balmer lines.  Figure~1 shows a montage of spectra around the
H$\alpha$ line.  Dashed lines indicate the wavelengths of H$\alpha$,
[N~II]~$\lambda\lambda$6548,6583, and [S~II]~$\lambda\lambda$6716,6730
for a best-fitting redshift of $0.21103 \pm 0.00013$.  Of particular
importance is the weak but clearly present [S~II] doublet at the same
redshift as H$\alpha$ and [N~II].  The [S~II] lines are so broad as to
be almost completely blended, but it is clear that all of the
emission-line peaks can be attributed to the narrow components, either
H$\alpha$ or forbidden lines, at a single redshift.  We also measured
the [O~II]~$\lambda$3727 redshift in the 1998 June spectrum, which
gave $z = 0.21096$, consistent with the other low-ionization lines.
We therefore adopt $z=0.21103 \pm 0.00013$ as the systemic
(low-ionization) redshift.

Next, we turn to the region around H$\beta$.  In Figure~2, we
draw dashed lines at the low-ionization redshift of 0.21103 to see
how well this agrees with [O~III] and H$\beta$.  The agreement with
[O~III] is very good.  Although [O~III] prefers a slightly
lower redshift of $0.21063 \pm 0.00016$, the corresponding difference
of only $\approx 100$~km~s$^{-1}$ between high- and low-ionization forbidden
lines is common.  This comparison assures us that we have correctly
identified the peaks on the H$\alpha$ line with their narrow-line region
components.  It is difficult to evaluate the agreement with narrow
H$\beta$, predicted to fall at 5887~\AA, because,
as we shall see, it nearly coincides with
with Na~I~D absorption in our Galaxy's interstellar medium (ISM).
In any case, it is clear that the large velocity widths of the
strong narrow emission lines, 
FWHM = 700~km~s$^{-1}$ and FWZI = 2500~km~s$^{-1}$ as exemplified
by [O~III], have confused previous interpretations of the broad 
Balmer-line profiles in OX~169.

\subsection{ISM Absorption Lines}

In Figure~2, we also draw dotted lines at the wavelengths
of Na~I~$\lambda\lambda$5890,5896 at zero velocity.
It is evident that Na~I~D
absorption is present near the peak of the broad H$\beta$ line, and
that this absorption is responsible for the peculiar appearance of
H$\beta$ that has been noted by previous authors.
The explanation of this feature as interstellar Na~I~D absorption
fits the data in striking detail.  The doublet is resolved in the
spectra of highest resolution, particularly from the KPNO 2.1m,
and the wavelengths of the pair fall exactly at zero velocity.
It is difficult to measure the absorption-line equivalent widths exactly
because they are superposed on the peak of an emission line which is a composite
of a narrow and broad component.  Our best estimate, fitting a Gaussian
to the peak of the broad emission line, yields an equivalent
width of about 0.6~\AA\ for each of the D$_1$ and D$_2$ lines. 

We do not regard any other explanations of this doublet as plausible.
For example, it is not reasonable to suppose that there
is a central reversal consisting of a narrow component of
H$\beta$ emission inside a self-absorbed or double-peaked broad emission
line.  This would require narrow H$\beta$ to be redshifted by
$\approx 300$~km~s$^{-1}$ with respect to narrow H$\alpha$ and the
low-ionization forbidden lines, all of which agree internally to an accuracy of
about 50~km~s$^{-1}$.
There is also no evidence that Na~I~D emission from the night sky
is contaminating these spectra.  The accuracy of our sky subtraction
is evidently very good, given the detailed reproducibility among the many
spectra in this region and the lack of any systematic problems with
other night-sky emission lines that are stronger than Na~I~D,
such as [O~I]~$\lambda$5577 and [O~I]~$\lambda$6300.  SF were also
careful to rule out the possibility of errors in sky subtraction
in their spectra, but they did not consider the possibility
of interstellar Na~I~D absorption which, of course, persists
after accurate subtraction of the night-sky emission.

Additional interstellar absorption lines whose equivalent widths are
known to correlate with that of Na~I~D are present in the spectrum of OX~169.
These are shown in Figures~3 and 4. The 1996 Lick spectrum covers the
region of Ca~II~H \& K ($\lambda\lambda$3968.5,3933.7), which have
equivalent widths of 0.14 \AA\ and 0.30 \AA, respectively.  As shown
by the correlations presented in Hobbs (1974), the Ca~II and Na~I~D
absorption-line strengths are consistent with the moderate H~I column
at these coordinates ($\ell,b = 72.\!^{\circ}116,-26.\!^{\circ}084$)
of $8.2 \times 10^{20}$~cm$^{-2}$ (Stark et
al. 1992), and with the extinction $E(B-V) = 0.111$ estimated from
{\it IRAS} 100~$\mu$m maps (Schlegel, Finkbeiner, \& Davis 1998).  In
the {\it HST\/} spectrum, the Mg~II~$\lambda\lambda$2795.5,2802.7
doublet is strong, with equivalent widths of 1.57 and 1.03 \AA,
respectively, and Fe~II, Mn~II, and Mg~I absorption lines are present
as well.  All of these features support the hypothesis that a modest
interstellar Na~I~D absorption is to blame for the
peculiar appearance the H$\beta$ emission line in OX~169.
 
\subsection{Broad Emission Lines}

The shapes of the broad emission lines in OX~169 are certainly quite
varied.  Figure~5 shows examples of all of the broad emission lines that
can be extracted from our optical spectra, as well as from the
{\it HST\/} spectra.  The chosen zero velocity point corresponds to the narrow-line 
redshift of 0.21103. It is interesting that narrow Ly$\alpha$ absorption
is present at exactly this redshift, which is presumably the systemic
redshift of the host galaxy.  The peak of Ly$\alpha$ emission, however,
is at $z = 0.21334$, which is redshifted by $\approx 570$~km~s$^{-1}$
from the narrow lines.
The other broad emission lines are also redshifted, by up to $\approx 700$
km~s$^{-1}$ as determined by Gaussian fits.  Table~2 lists all of the
broad emission-line widths and shifts from the spectra illustrated in
Figure~5.
In addition to this first-order characterization, there are complicating
features in some of the line profiles, such as an extended
red wing on H$\beta$ which may be due in part to
Fe~II~$\lambda\lambda$4923,5018.  Supporting this interpretation are
the appearance of broad Fe~II multiplets around $4570$~\AA\ and $5250$~\AA,
and the fact that such an extended wing is not present in H$\alpha$.
H$\gamma$ may contain a weak contribution from [O~III]~$\lambda$4363,
which could have contributed to the double-peaked
appearance of this line in previous studies.
Broad wings are present on Ly$\alpha$, but its red wing may
also contain a contribution from N~V~$\lambda$1240.

The C~IV and C~III] line profiles are noisy, and difficult to characterize.
Partly inspired by the previous reports of double-peaked Balmer
lines in OX~169, Marziani et al. (1996) wrote that the C~IV line is
probably double-peaked, although the evidence was not strong in their
view.  We also note that the apparent
associated C~IV absorption feature at $-500$~km~s$^{-1}$ is not highly
significant, and it may be an
intrumental artifact because it is narrower than the resolution.
The C~III]~$\lambda$1909 line is difficult to interpret because it is
likely that a contribution from Si~III]~$\lambda$1892 is present,
as well as lines of the Fe~III multiplet UV34.  Aoki \& Yoshida (1998) and
Wills et al. (1999) show that Si~III]~$\lambda$1892 
typically contributes 20--30\% of the flux in this blend, and there
is some evidence for such a contribution in Figure~5.

In summary, there are small shifts between the broad-line and
narrow-line velocities, as well as differences among the broad-line
profiles themselves.  These effects are well known among quasars.
However, there is no evidence for a
two-component broad-line region, or self-absorption in any of the
non-resonance lines.  After 10 years of monitoring OX~169, we consider
that a double-peaked description of its broad emission lines is pretty
much ruled out.

\section{Discussion}

As it turns out, previous descriptions of the spectrum
of OX~169 each had some element of truth, but were misled by one or more
effects. There {\it is} absorption in the H$\beta$ line (Smith 1980),
but it is caused by Na~I in the Galaxy's ISM.  The BLR {\it is} redshifted
from the narrow-lines, not by 1200 km~s$^{-1}$ as claimed by Gaskell (1981),
but by 300--700 km~s$^{-1}$.   SF rejected
both of these scenarios, and discussed either a double-peaked
accretion disk line profile, or a binary BLR.
The principal argument that SF developed in
favor of a binary BLR employed line-profile variability,
which they interpreted as independent behavior of a blueshifted
and redshifted line component.  Although our own spectra do
not support such a description, the spectra of SF,
particularly before 1989, remain as
evidence that is independent of
ours.  Therefore, for completeness, we explore the arguments
by which their spectra were interpreted, and review
what has been learned about variability of double-peaked
emitters in the past decade.

\subsection{Tests of the Binary BLR Hypothesis}

Actually, the variability studied by SF was limited to just
two spectra of H$\beta$, taken in 1983 and 1989.  The assumption
behind the analysis of SF is that, when a line shape changes,
the profile can be uniquely decomposed by differencing into two
components, a variable part and a constant part.
The resulting pair of line profiles were in turn attributed
to two spatially separated sources.  In our opinion, it is
doubtful that a reliable interpretation of variability
can be extracted from just two spectra.  In this particular case,
it is also important to evaluate the assumptions behind the method.
As SF state, for their procedure to have meaning, the light-travel
time across the broad-line region must be short compared to the
time scale of variability of the photoionizing continuum.
A pair of additional requirements that were {\it not} stated are
1) that each variable photoionizing source does not affect the other's emission-line
region, and 2) that variability of a photoionizing source
is the {\it only} mechanism of line profile variability.  But all of
these requirements together amount to {\it assuming} most of the
properties of the desired solution, namely, that
a pair of photoionizing sources are associated with spatially distinct
BLRs which have stationary velocity fields and are immune from the 
effects of each other's radiation.  It does not seem possible that
a single difference spectrum could be used to justify all of the
required assumptions without employing a circular argument.

A number of intensive monitoring programs have been conducted over
the past decade which bear upon these issues.  First, a sensitive
search for the smoking gun of the binary BLR model in  
three bona fide double-peaked emitters
yielded interesting but null results (Eracleous et al. 1997).
The absence of long-term, systematic velocity variations
characteristic of a double-lined spectroscopic
binary effectively ruled out the binary BLR model for all reasonable
black hole masses in Arp~102B, 3C~390.3, and 3C~332.  The factor which
makes this test feasible in a reasonable period of time is a large
velocity separation of the emission-line peaks.
The observed {\it absence} of radial velocity variations can be translated into a lower limit on the mass of 
the assumed binary,
$$M\ >\ 4.7 \times 10^8\,(1+q)^3\,\left({P \over 100\,{\rm yr}}\right)\,
\left({v_1\,{\rm sin}\,i \over 5000\ {\rm km\,s^{-1}}}\right)^3\ M_{\odot}, 
\eqno(1)$$
where $M = M_1 + M_2$, $q = M_1/M_2$, $P$ is an observed lower limit
on the orbital period,
and $v_1\,{\rm sin}\,i$ is the observed radial velocity of $M_1$.
Since the mass depends on the cube of the velocity, those
line profiles with peaks that are displaced by 5,000~km~s$^{-1}$
or more can provide a very sensitive test of the hypothesis
in a couple of decades.  Eracleous et al. (1997) eliminated all binary
masses less than $10^{10}\,M_{\odot}$ in Arp~102B, 3C~390.3, and 3C~332.
A previous analysis of Gaskell (1996), which found 
striking evidence for a
radial velocity drift and thus binary orbital motion in 3C~390.3,
was contradicted by the longer time span of the observations
made by Eracleous et al. (1997) in which the
trend did not continue as expected for a spectroscopic binary.

This demonstrated absence of binary BLRs in those three objects
implies that there must be a mechanism by which a single black hole
can produce a double-peaked emission line, but it does not rule
out a scenario in which an unseen black hole perturbs the emission-line
velocity of another.  However, as can be
seen by inverting equation (1) for the orbital period $P$,
it might be difficult to discover such a single-lined spectroscopic
binary in an object like OX~169 for which the radial velocity displacement
of the broad emission lines is $\leq 700$~km~s$^{-1}$.

\subsection{Line-Profile Variability: Dynamics, not Reverberation!}

A second major development in the study of emission-line variability
is the realization that line-profile variability is not the result of
light-echo effects, i.e., reverberation.  All of the Seyfert monitoring
campaigns have shown that, although the total {\it intensity} of an emission
line is modulated in response to the intensity of the ionizing continuum
with a lag of days to months, the {\it shape} of the line changes hardly,
if at all, on these time scales (Ulrich 1991; Wanders \& Peterson 1996;
Kassebaum et al. 1996).  In particular, both sides of the double-peaked
emission line in 3C~390.3 respond simultaneously to continuum
variations (Dietrich et al. 1998; O'Brien et al. 1998).
On the other hand, we have learned
that major changes in line shapes on long time scales of years to 
decades is ubiquitous, especially in double-peaked emitters
(Veilleux \& Zheng 1991; Newman et al. 1997;
Storchi-Bergmann et al. 1995; Gilbert et al. 1998), but that these 
slow profile variations are not responses to changes
in the ionizing continuum.  Rather, they must be due to physical changes
in the velocity field of the emitting gas, i.e., {\it dynamical motions}.
Some of the most dramatic examples are found in the emergence of new
double-peaked broad emission lines in well known objects which had no
such component in the past, such as Pictor~A (Halpern \& Eracleous 1994;
Sulentic et al. 1995), M81 (Bower et al. 1996), and NGC~1097 (Storchi-Bergmann, Baldwin, \& Wilson 1993).

Much of the recent effort in modeling line-profile variability has
focussed on dynamical motions such as hot spots and spiral waves in
accretion disks
(Zheng, Veilleux, \& Grandi 1991; Chakrabarti \& Wiita 1994;
Newman et al. 1997; Gilbert et al. 1999),
tidal disruption of stars, and precessing eccentric accretion disks 
(Eracleous et al. 1995; Storchi-Bergmann et al. 1997).  This is not to
say that a universally applicable model of line profile variability
is in the offing.  On the contrary, it is a warning that one should not
expect to extract a dynamical model of a quasar broad-line region from
two snapshots of an emission-line profile, or even from a dozen.
In all of these studies, the double-peaked line profile is treated 
as a dynamic whole to be modeled with an evolving velocity field,
as there is evidently no simple decomposition of the profile
into a pair of independent, stationary entities.

Since much recent modeling involves an accretion-disk origin for
the emission-lines, we wish to address here a stock criticism of the
accretion-disk hypothesis which seems to persist, unjustifiably in
our opinion, and should be put to rest.
Several authors have noted that {\it if\/}
line-profile variability is caused by the response of an
axisymmetric accretion disk to a variable central photoionizing source,
then the blue and red sides of an emission-line profile should vary
in concert (e.g., Miller \& Peterson 1990).
While this is a valid proposition as far as it goes, recent references 
to it have
omitted the clause beginning with the word {\it if}, asserting, in effect,
that line profile variability {\it is} caused by reverberation, and that 
it must therefore be symmetric in a disk model (e.g., Gaskell 1996).
Since line-profile variability is generally observed to be
asymmetric, disks are disfavored according to this argument.
But as reviewed here, there is now ample evidence that observed
line-profile variability has a much longer
time scale than can be explained by reverberation, and must
therefore be a dynamical effect.
On {\it short} time scales, line peaks {\it do} vary in concert,
and are thus consistent with a disk-like velocity field.
While reprocessing undoubtedly
plays an important role in explaining the spectra of quasars,
we would do well to remember that not all variability must or can be
attributed to reprocessing, and this appears especially to be the case
for emission-line profiles.

\subsection{Accretion-Disk Emission Revisited}

The question then arises, is there anything of general interest
to be learned from the broad emission lines in OX~169?  Having
patiently collected data for 10 years, we feel the obligation to
engage in at least some speculation of our own.
Since we have argued that the principal characteristics of the broad
lines are 1) a slight redshift with respect to the narrow lines, and
2) differences among their widths, it would appear that the most
natural location for their origin would be on the surface of an accretion
disk (the second-best hypothesis according to SF). 
The redshifts can be
understood as the combined effect of gravitational and transverse
redshift on circular motion viewed close to the rotation axis.
The net redshift $\Delta v$ in the weak-field limit is simply
$\Delta v/c \approx (3/2)\,(r_{\rm g}/r)$, where $r$ is the orbital radius
and $r_{\rm g}$ is the gravitational radius $GM/c^2$.  The line
width $v$ is dominated by the longitudinal Doppler shift, such that
$v/c \approx ({r_g/r})^{1/2}\ {\rm sin}\,i$.  For example, since we measure
for the H$\gamma$ line in OX~169 $\Delta v \approx 730$~km~s$^{-1}$ and
$v \approx 4500$~km~s$^{-1}$, we estimate crudely
that $r \approx 620\,r_{\rm g}$ and $i \approx 22^{\circ}$.
Such mild relativistic effects have
long been hypothesized to explain such asymmetries that are
generally found in emission-line profiles (e.g., Corbin 1997).

As mentioned by SF, the photoionization models of Dumont \& Collin-Souffrin (1990a,b), in which a central photoionizing source illuminates the outer
accretion disk, 
naturally produce lines of different widths because the physical conditions
vary with radius in the disk.  In these calculations, H$\alpha$ is narrower
than H$\beta$ because the lower-order transitions saturate first at
high flux levels in the inner disk.  Another successful
prediction of the model is that Mg~II is narrower than H$\alpha$.
Mg~II is preferentially produced in the outer disk where, at low flux levels,
it is enhanced over H$\alpha$ because the lines are formed by recombination
and not by collisional excitation.  Thus, the trend among the
emission-line widths in OX~169 is
largely in accord with the disk photoionization model.
In addition, the accretion-disk
wind model (Chiang \& Murray 1996; Murray \& Chiang 1997,1998)
may well be applicable, and will have additional affects on the emission-line profiles.
 
Although SF considered the accretion-disk hypothesis in light of a presumed
double-peaked emission line, there is no requirement that disk lines
be double peaked, as numerous authors, including
Dumont \& Collin-Souffrin (1990a,b), Jackson, Penston, \& P\'erez (1991),
and Murray \& Chiang (1997) have explained.
The most extreme of the observed double-peaked emitters may be easiest
to {\it recognize} as disk-like because of the wide separation of their
peaks, but this may be the exceptional case
that obtains when the ratio of outer to inner
radius is small, i.e, only $\sim 3$.  As the outer radius of the line
emitting region increases, the two peaks merge together at small velocity,
which may be the more general rule.  Radiative transfer effects in the
lines also tend to make single-peaked profiles.
The small inclination inferred for
OX~169 would be consistent with its single-peaked line profiles, and
also with its core-dominated radio source.

\section{Conclusions and Future Work}

We have shown that the Balmer lines in OX~169
are neither self-absorbed nor double peaked.  All previous analyses
of its spectra were led astray by some combination of
the following effects: 1) The forbidden lines of OX~169
are unusually strong and broad, consequently,
[N~II]~$\lambda$6583 masquerades as an additional component
of H$\alpha$.  2) H$\beta$ coincidences with Galactic Na~I~D absorption,
which has an equivalent width similar to the spurious ``trough'' between
H$\alpha$ and [N~II]~$\lambda$6583.  3)  The broad emission lines are
redshifted by as much as 700~km~s$^{-1}$ from the forbidden lines.
In spectra obtained over the past decade,
we see no substantial change in the character of the line profiles,
and no indication of intrinsic double-peaked structure once the above effects
are recognized.  In support of this interpretation, we show that 1)
the Na~I~D doublet is resolved in absorption, 2) ISM absorption in Ca~II H
and K and Mg~II are detected at a strength consistent with that of
Na~I~D, 3) the Mg~II, Ly$\alpha$, and
H~$\gamma$ emission lines are single peaked, and 4) all of the emission-line
redshifts are consistent once they are correctly
attributed to their permitted and forbidden-line identifications.

A systematic shift of up to 700 km~s$^{-1}$ between broad
and narrow lines is seen, but such differences are common and could
be due to gravitational and transverse redshift in a disk-like broad-line
region viewed at small inclination.
The single peaked nature of the emission lines is not an
obstacle to a disk model, and may in fact be the general rule, while
double-peaked lines are the exception.  Long-term variability of the
emission-line profiles in OX~169 appears to be modest and unexceptional,
and is probably due to dynamical motions.  Ultimately, our understanding
of why quasars vary will have to involve dynamics.

Stockton \& Farnham (1991) interpreted the line profiles of OX~169 in
terms of a binary BLR, which was especially intriguing since they also
found an apparent tidal tail in the host galaxy, and speculated that a
recent merger had supplied the nucleus with a pair of black holes
which was now coalescing.  Strictly speaking, our revised description
of the line profiles is not to be taken as evidence against the {\it
presence} of a binary black hole, but only for the absence of two
separate emission-line regions.  In view of the mounting evidence for
the ubiquity of black holes in galactic nuclei, the formation of
binary black holes in galaxy mergers should be relatively common.
Such binaries could spend anywhere from $10^8 - 10^{10}$~yr at
separations of $0.01 - 0.1$~pc (Begelman et. al 1980), during which
their orbital motion might be detected.  According to equation (1), it
would be worthwhile to monitor emission lines that have peaks
displaced by more than 1500~km~s$^{-1}$ for evidence of binary motion,
especially if such displacements are not easily compatible with
gravitational redshift alone, e.g., if they are blueshifted.
We have several such candidates under surveillance.
Even if only one emission-line region
exists in such a system, the orbital acceleration by the unseen
black hole could perturb the emission-line velocities in the manner of
a single-lined spectroscopic binary.  Such black hole binaries of $M
\approx 10^8\,M_{\odot}$ would undergo detectable orbital motion in
just a couple of decades.  If these candidates were also compact
VLBI radio sources, it would feasible to obtain direct confirmation
via proper motion of order microarcseconds per year (Eracleous et al. 1997).
Indirect evidence for binary orbital motion may be present
in the wiggles of a milliarcsecond radio jet
(Kaastra \& Roos 1992; Roos, Kaastra, \& Hummel 1993).

Unfortunately, OX~169 is no longer a prime candidate
for such a monitoring program.  Over the past decade its
emission lines have revealed little evidence for unusual
velocities, and no other peculiarities that inspire thoughts of binarity.
Regretfully we opine, this is the end of the lines for OX~169.

\acknowledgements 

This work was based in part on observations made with the NASA/ESA
Hubble Space Telescope, obtained from the data archive at the Space
Telescope Science Institute, which is operated by the Association of
Universities for Research in Astronomy, Inc. under NASA contract 
NAS~5-26555.
 
\clearpage

%
%

\clearpage

\figcaption[halpha.ps]{Spectra around the H$\alpha$ line of
OX~169 which have been renormalized and shifted vertically
for clarity. The dashed lines correspond to the wavelengths expected
for [N~II]~$\lambda$6548, H$\alpha$, [N~II]~$\lambda$6583,
[S~II]~$\lambda$6716, and [S~II]~$\lambda$6730, all
at $z = 0.21103$. \label{fig1}}

\figcaption[hbeta.ps]{Spectra around the H$\beta$ line of
OX~169 which have been normalized to the flux of [O~III]$\lambda$5007
and shifted vertically for clarity. The dashed lines correspond 
to the wavelengths expected for H$\beta$ and 
[O~III]~$\lambda\lambda$4959,5007 at $z = 0.21103$.
Dotted lines mark the wavelengths of the Na~I~D doublet
at $z = 0$. \label{fig2}}

\figcaption[fig_4.ps]{Ca~II H and K absorption from 
the ISM in the spectrum of OX~169.  The H and K equivalent widths are
0.14 \AA, and 0.30 \AA, respectively. \label{fig3}}

\figcaption[fig_3.ps]{Galactic ISM absorption lines in
the {\it HST\/} ultraviolet spectrum of OX~169.  The expected
zero-velocity wavelengths of various absorption lines are
indicated, as are their relative oscillator strengths by
the vertical length of the tick mark. The equivalent
widths of Mg~II~$\lambda\lambda$2795.5,2802.7 are 1.57 \AA, and
1.03 \AA, respectively. \label{fig4}}

\figcaption[mce.ps]{Broad emission line profiles from the
optical and UV spectra of OX~169.  Continuum has been
subtracted. Zero velocity in this figure
corresponds to the narrow-line redshift $z = 0.21103$ . \label{fig5}}

\begin{deluxetable}{llccc}
\footnotesize
\tablecaption{Optical and UV Spectroscopy of OX 169. \label{tbl-1}}
\tablewidth{0pt}
\tablehead{
\colhead{UT Date} & \colhead{Telescope}   & \colhead{Exposure time}   & 
\colhead{Wavelength range} &  \colhead{Resolution} \nl
 & & (s) & (\AA) & (\AA)} 
\startdata
1989 Nov. 7  & MDM 2.4m   &  1800  &  $5300-8100$  &  12  \nl
1990 May 30  & KPNO 2.1m  &  2700  &  $6300-8300$  &   7  \nl
1992 Jan. 7  & {\it HST\/} FOS    &  2043  &  $1160-1600$  &   1  \nl
1992 Jan. 7  & {\it HST\/} FOS    &  1001  &  $1600-2310$  &   1.5   \nl
1992 Jan. 7  & {\it HST\/} FOS    &  1047  &  $2220-3280$  &   2  \nl
1993 Dec 13  & KPNO 4m    &  1800  &  $5000-9800$  &   8  \nl
1994 July 4  & KPNO 2.1m  &  4000  &  $5500-8500$  &   4  \nl
1995 June 4  & KPNO 2.1m  &  4000  &  $5500-8500$  &   4  \nl
1996 June 15 & KPNO 2.1m  &  3915  &  $5500-8500$  &   4  \nl
1996 Oct. 11 & Lick 3m    &  4800  &  $3200-4400$  &   5  \nl
1996 Oct. 11 & Lick 3m    &  4800  &  $5600-8300$  &   5  \nl  
1997 June 9  & KPNO 2.1m  &  3600  &  $5500-8500$  &   4  \nl
1997 Sep. 29 & KPNO 2.1m  &  3600  &  $5500-8500$  &   4  \nl
1998 June 27 & MDM  2.4m  &  3600  &  $6300-8300$  &   5  \nl
1998 June 29 & MDM  2.4m  &  2500  &  $4400-6400$  &   7  \nl
\enddata
\tablenotetext{}{} 
\end{deluxetable}

\begin{deluxetable}{lccrc}
\footnotesize
\tablecaption{Emission-Line Widths and Shifts. \label{tbl-2}}
\tablewidth{0pt}
\tablehead{
\colhead{Line Identification} & \colhead{Rest}  & \colhead{Measured}  & 
\colhead{Shift$^a$} &  \colhead{FWHM} \nl
 & Wavelength (\AA) & Wavelength (\AA) & (km s$^{-1}$) & (km s$^{-1}$) }
\startdata
H$\alpha$   &  6562.79  &  7956.12  &  320  &  3770  \nl
H$\beta$    &  4861.33  &  5895.88  &  440  &  4550  \nl
H$\gamma$   &  4340.46  &  5269.20  &  730  &  4490  \nl
Mg~II       &  2799.07  &  3397.10  &  650  &  2650  \nl
C~III]      &  1908.73  &  2312.18  &   80  &  5380  \nl
C~IV        &  1549.48  &  1879.67  &   30  &  6930  \nl
Ly$\alpha$  &  1215.67  &  1475.02  &  570  &  5030  \nl
\enddata
\tablenotetext{}{$^a$ Velocity with respect to the systemic
redshift of $z = 0.21103$.  Velocities are all positive, therefore 
redshifted.} 
\end{deluxetable}

 
\plotone{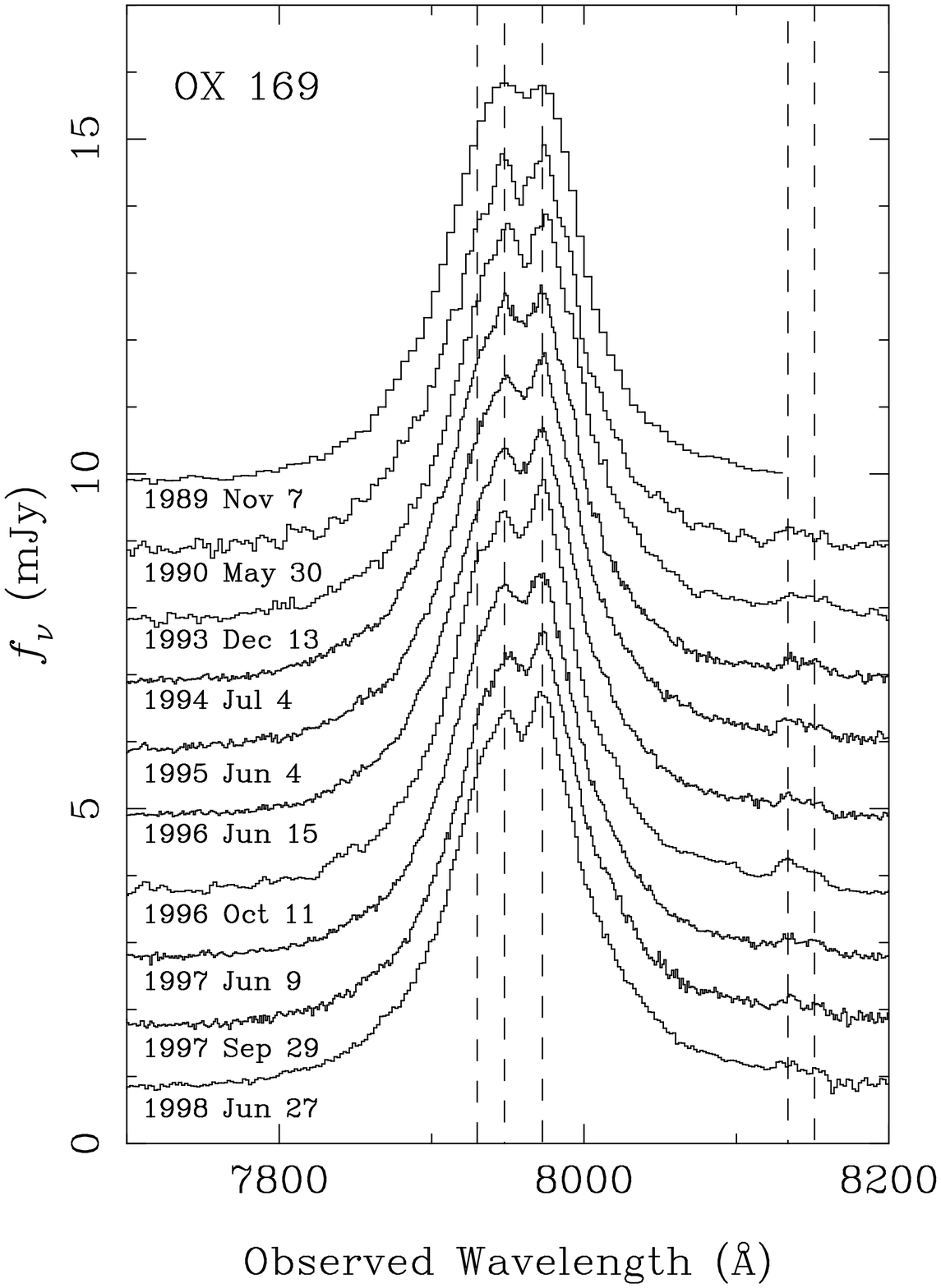}
 
\plotone{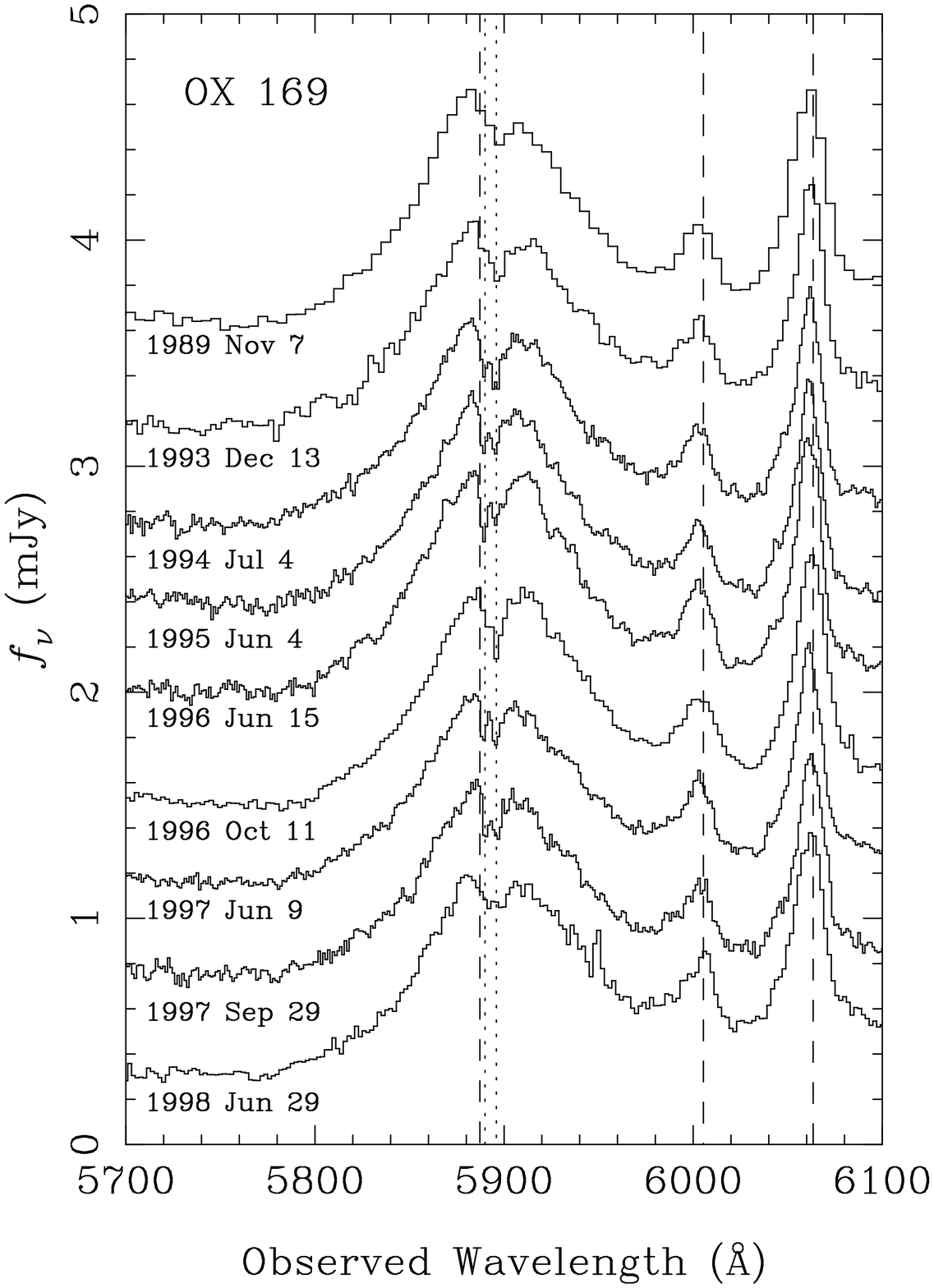}

\epsscale{0.5}
\plotone{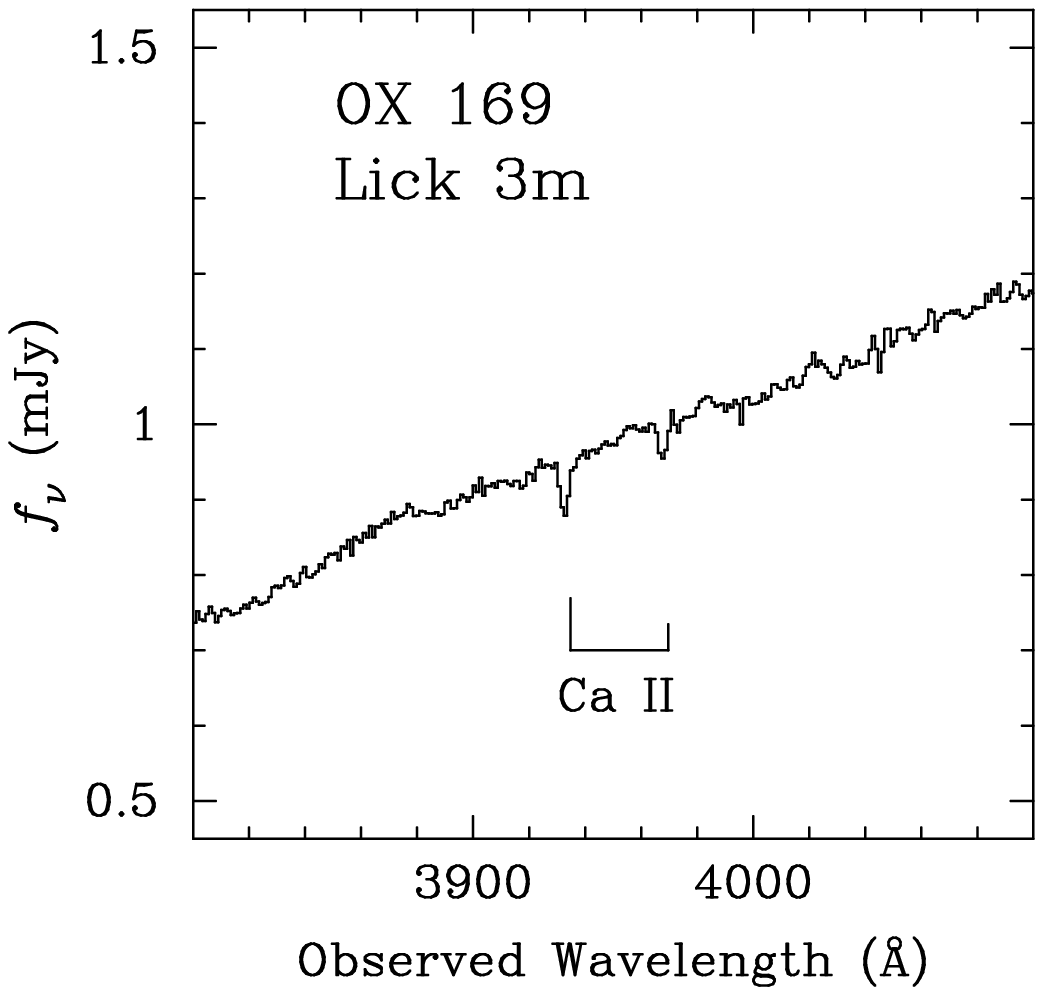}

\plotone{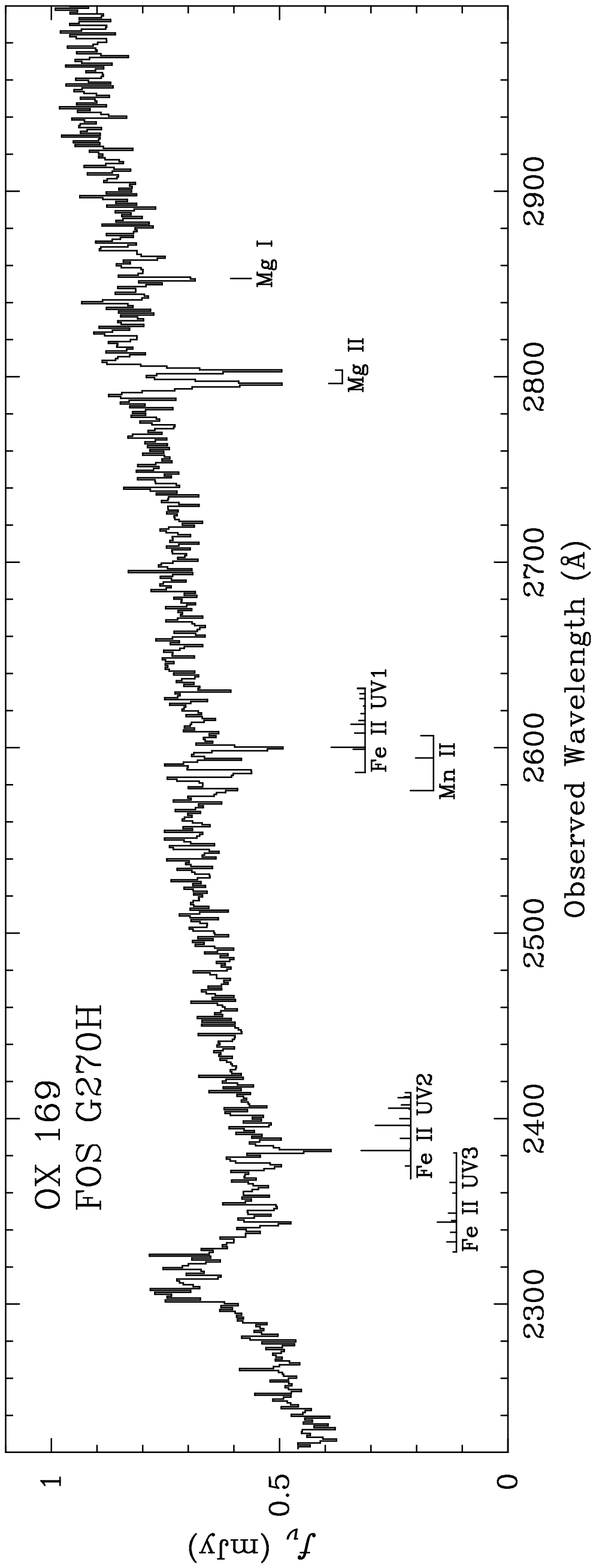}

\epsscale{0.65}
\plotone{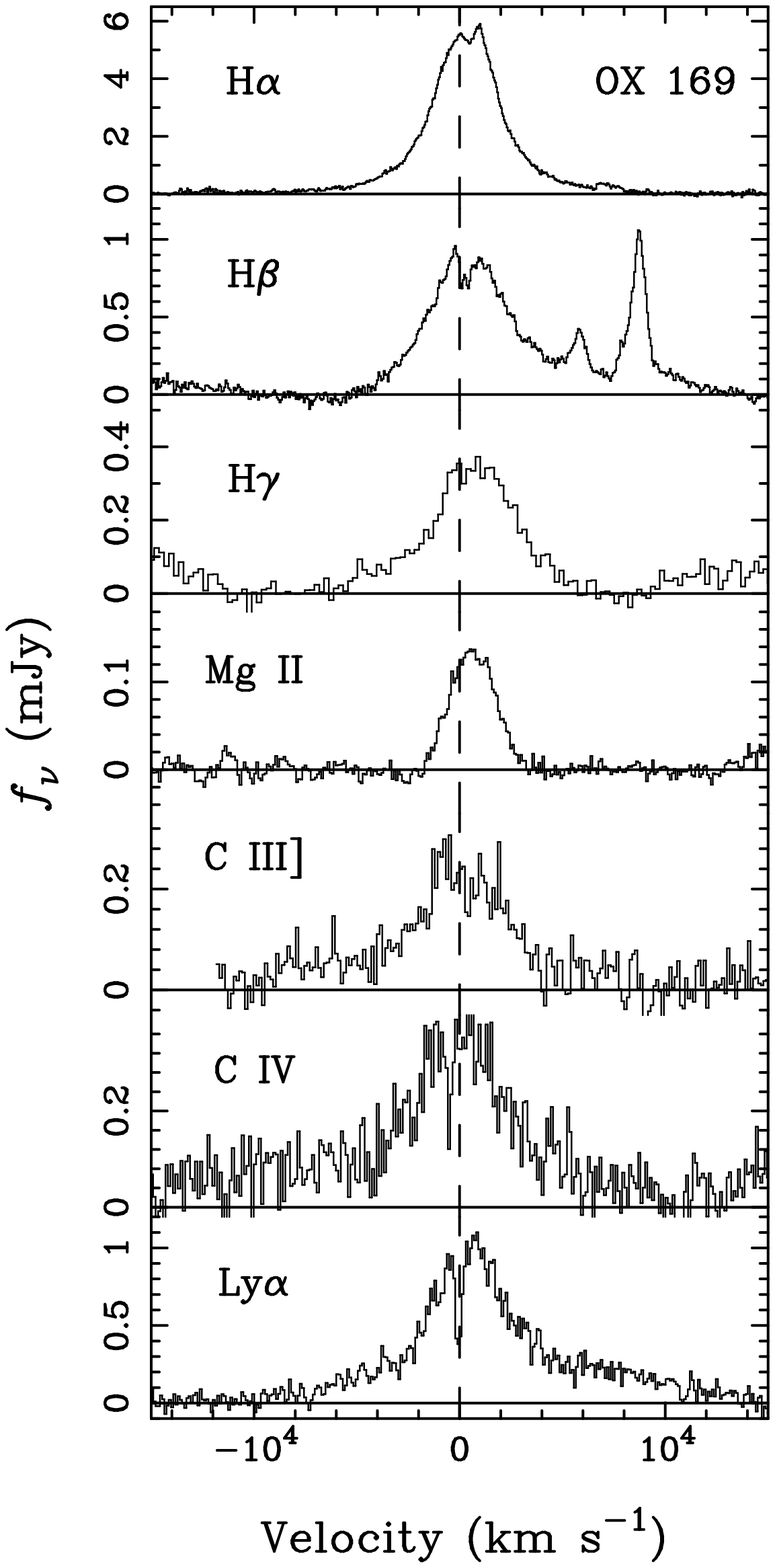}

\end{document}